\documentclass{phyeauth}
\usepackage{graphicx}
\usepackage{amsmath}
\usepackage{amssymb}

\begin{document}


\begin{frontmatter}

\title{Probing the Electrostatics of Integer Quantum Hall Edges with Momentum-Resolved Tunnel Spectroscopy
 }

\author[WSI]{M.~Huber},
\author[WSI]{M.~Grayson\thanksref{thank1}},
\author[WSI]{D.~Schuh},
\author[WSI]{M.~Bichler},
\author[WMI]{W.~Biberacher},
\author[Regensburg]{W.~Wegscheider},
and
\author[WSI]{G.~Abstreiter}

\address[WSI]{Walter Schottky Institut, Am Coulombwall, 85748 Garching, Germany}

\address[WMI]{Walther-Meissner-Institut, Walther-Meissner-Strasse 8, 85748 Garching, Germany}

\address[Regensburg]{Universit\"at Regensburg, Universit\"atsstrasse 1, 93040 Regensburg, Germany}

\thanks[thank1]{
Corresponding author.\\E-mail: mgrayson@alumni.princeton.edu}

\begin{abstract}
We present measurements of momentum-resolved magneto-tunneling
from a perpendicular two-dimensional (2D) contact into integer
quantum Hall (QH) edges at a sharp edge potential created by
cleaved edge overgrowth. Resonances in the tunnel conductance
correspond to coincidences of electronic states of the QH edge and
the 2D contact in energy-momentum space. With this dispersion
relation reflecting the potential distribution at the edge we can
directly measure the band bending at our cleaved edge under the
influence of an external voltage bias. At finite bias we observe
significant deviations from the flat-band condition in agreement
with self-consistent calculations of the edge potential.
\end{abstract}

\end{frontmatter}

\section{Introduction}
The quantum Hall (QH) effect arises due to energy gaps developing
in the spectrum of two-dimensional electron systems with a
perpendicular magnetic field. At specific ratios of electron sheet
density and magnetic field, when the Fermi-energy is in such a
gap, the only low-energy excitations in a finite quantum Hall
sample are located at the boundary, where the energetically bent
Landau-levels intersect the Fermi-energy. These states at the
sample edge exhibit one-dimensional (1D) chiral transport
behavior, i.e. quantized conductance, independent of the exact
electrostatics at the sample boundary. Typical transport
measurements therefore do not provide information about the edge
electrostatics (e.g. depletion lengths or edge reconstructions) or
the exact spectrum of the edge excitations. Therefore different
techniques like tunneling into the QH edge were introduced to
study electron correlation in QH edges.

Due to the atomic precision of confinement potentials and tunnel
barriers, the method of cleaved-edge overgrowth (CEO) \cite{CEO}
is of particular interest for fabricating low dimensional tunnel
structures. 2D-3D tunneling at a sharp edge was used to study the
density of states of fractional quantum Hall edges \cite{Chang PRL
96}. In such a geometry the fingerprint of non-Fermi-liquid
behaviour was power law behaviour in the tunneling density of
states over a continuum of fractional quantum Hall edge channels
\cite{Grayson PRL 98}. Subsequent experiments showed evidence for
plateau structure at $\nu =1/3$ \cite{Chang PRL 01} as well as a
drastically shifted power law plot \cite{Hilke} indicating sample
dependence in the observed characteristics. In this context it was
suggested that the exact potential shape at the edge might be
crucial for defining the correlations at the QH edge. In this
paper we present a new geometry where the 3D tunnel contact is
replaced by a high mobility 2D system \cite{Huber Physica E}. This
allows momentum resolved tunnel spectrocopy of the QH edge. In the
fractional QH regime this geometry is predicted to probe the
spectral function of charged and neutral modes at specific filling
factors \cite{Zulicke}. In this paper we study the edge dispersion
of Landau levels in the integer QH regime and deduce information
about the potential distribution at the edge. In contrast to the
geometry presented by W. Kang, et al. \cite{Kang nature}, where
tunneling between two quantum Hall edges is studied, in our sample
the contact probing the quantum Hall edge is a simple Fermi system
whose properties are not affected by the magnetic field.

\section{Sample design}
\begin{figure}[h]
\begin{center}\leavevmode
\includegraphics[width=0.7\linewidth]{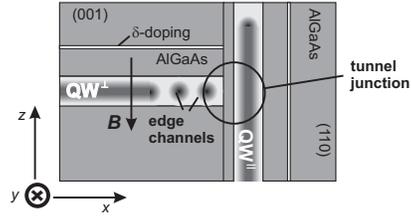}
\caption{The samples are fabricated by cleaved edge overgrowth.
Two quantum wells ($QW^{\perp}$ and $QW^{\parallel}$) are arranged
in a T-shape. A magnetic field $B$ creates quantum Hall edge
states close to the tunnel
barrier.}\label{sample}\end{center}\end{figure}

The samples consist of two separately contacted perpendicular high
mobility quantum wells ($QW^{\perp}$ and $QW^{\parallel}$) forming
a T-shaped structure (Fig. \ref{sample}), where $^{\perp}$ and
$^{\parallel}$ are defined relative to the quantizing $B$ field.
These QWs consist of GaAs embedded in Al$_{0.32}$Ga$_{0.78}$As.
Using cleaved-edge overgrowth (CEO), a 150~\AA\ thick
(001)-quantum well ($QW^{\perp}$) is cleaved along the
perpendicular (110)-plane and overgrown with a 200~\AA\ thick
(110)-quantum well ($QW^{\parallel}$) in a second epitaxial growth
step. The quantum wells are separated from each other by a 50~\AA\
thick 320~meV high Al$_{0.32}$Ga$_{0.78}$As tunnel barrier. Both
QWs are modulation doped with a Si-$\delta$~layer 500~\AA\ and
400~\AA\ away from the respective QWs. The electron sheet density
in the bulk of $QW^{\perp}$ and $QW^{\parallel}$ after
illumination is $2\times 10^{11}~{\rm cm}^{-2}$ for both and they
are 5000~\AA\ and 3600~\AA\ below the surface, respectively. In
the $y$-direction the sample geometry is translationally invariant
and the tunnel junction extends about 20~$\mu$m in width.

\section{Experimental Results}
\begin{figure}[h]
\begin{center}\leavevmode
\includegraphics[width=1.0\linewidth]{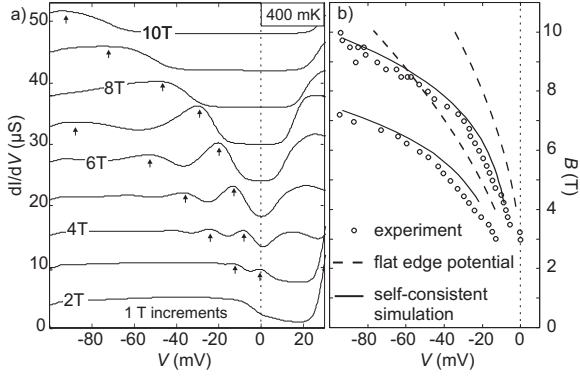}
\caption{a) Differential conductance $dI/dV$ plotted against
voltage bias ($V$) between $QW^{\perp}$ and $QW^{\parallel}$, with
successive traces shifted by $+6~\mu$S. The resonances due to
tunneling into the lowest two Landau levels are marked by arrows.
b) Positions of $dI/dV$ maxima for the lowest two Landau levels
plotted in the $V-B$ plane: The experimental results (circles)
differ significantly from the values determined from a flat edge
potential (dashed line) but are well described by a
self-consistently calculated edge potential (solid
line).}\label{cond}\end{center}\end{figure}

For our measurement we apply a magnetic field $B$ in the
$z$-direction perpendicular to $QW^{\perp}$ where it causes Landau
quantization and the formation of edge channels close to the
tunnel junction (schematically shown in Fig.~\ref{sample}).
Applying a voltage bias $V$ to $QW^{\parallel}$ with $QW^{\perp}$
grounded, we measure the differential tunnel conductance
$G~=~dI/dV$ in a $^3$He cryostat at temperatures of about 400~mK.
Fig.~\ref{cond} shows several conductance traces plotted against
applied voltage bias for a series of $B$-values between 2~T and
10~T. At $B$ fields above 2.5~T we observe clear maxima and minima
at low negative voltages. They become more pronounced and shift
towards negative bias at higher magnetic fields, with their
separation increasing with magnetic field from about 10~mV at
3.5~T to more than 50~mV at 7~T. The conductance at zero bias
disappears beyond a $B$-field of 4~T and the voltage range of
suppressed conductance around zero bias increases with higher
$B$-fields. At 10~T the conductance is suppressed down to almost
$-60$~mV.

\section{Discussion}
\begin{figure}[h]
\begin{center}\leavevmode
\includegraphics[width=0.95\linewidth]{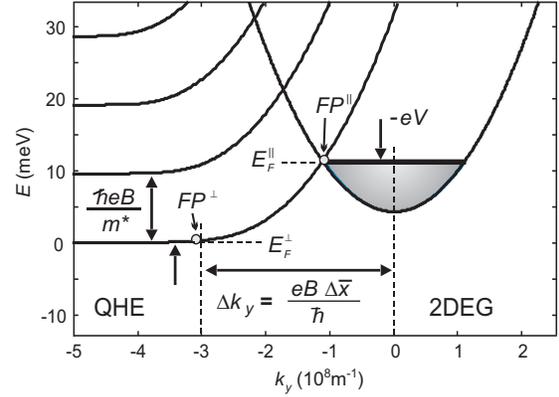}
\caption{Calculated dispersion relations $E(k_y)$ for the quantum
Hall system in $QW^{\perp}$ (QHE) and the 2DEG in $QW^{\parallel}$
according to Eq.~(\ref{schroedinger2}) and shown with a finite
bias $V$ between the two systems assuming a hard-wall potential.
The Lorentz force proportional to $B$ shifts the dispersion in
$k$-space and the bias $V$ shifts the
energy.}\label{schematic}\end{center}\end{figure}

In the presence of a magnetic field the electronic states in the
$QW^{\perp}$ are quantized to Landau levels with energy gaps
proportional to $B$. With an applied junction bias $V$ the
confining edge potential $\Phi (x,V)$ leads to a dispersion of the
Landau level energy $E_n^{\perp}(k_y)$ (see Fig. \ref{schematic},
left) in the vicinity of the tunnel barrier through the
Schroedinger equation

\begin{equation}
     \left[\frac{(\vec{p}-e\vec{A})^2}{2m^*}+\Phi (x,V)\right]\Psi _n(x,y)=E_n^{\perp}\Psi
     _n(x,y)~.
     \label{schroedinger}
\end{equation}

\noindent With a Landau gauge $\vec{A}(x,y)=xB\vec{\hat{y}}$,
where $x=0$ in the center of $QW^{\parallel}$, and taking
advantage of the translational invariance in $y$-direction by
expressing $\Psi _n(x,y)=\psi (x)e^{ik_yy}$, we can solve for the
motion in $x$:

\begin{eqnarray}
     \left[\frac{\vec{p}_x^2}{2m^*}+\frac{1}{2}\omega _c^2(x-\bar{x})^2+\Phi (x,V)\right]\psi
     _n(x) \nonumber
     \\
     =E_n^{\perp}(k_y)\psi _n(x)
     \label{schroedinger2}
\end{eqnarray}

\noindent where $\bar{x}=k_yl_B^2$ is the electron orbit guiding
center and $l_B=\sqrt{\hbar/eB}$ defines the magnetic length.
Fig.~\ref{schematic} shows the calculated $E_n^{\perp}(k_y)$
assuming a simple step function edge potential.




Alternatively the $k$-space dispersion $E^{\parallel}(k_y)$ of
$QW^{\parallel}$ exhibits a parabolic shape with a well defined
Fermi point $FP^{\parallel}$. From our choice of Landau gauge, the
mass parabola will always be centered at $k_y=0$, as shown in
Fig.~\ref{schematic}. Since Zeemann splitting in GaAs is small we
neglect the influence of the in-plane magnetic field on
$QW^{\parallel}$.

The translational invariance of the geometry in $y$-direction
together with the high mobility of the 2DEGs and the high
uniformity of the tunnel barrier causes both momentum $k_y$ and
energy $E$ to be conserved during tunneling. Graphically, this
means that tunneling is only allowed where the dispersion curves
intersect. Resonances in the tunnel conductance correspond to
coincidences of one of the quasi-Fermi-points ($FP^{\perp},
FP^{\parallel}$) with such a crossing point. Applying a voltage
bias the dispersion curves are shifted in energy, and the magnetic
field shifts them in momentum space with respect to each other
through the Lorentz impulse acquired by tunneling the effective
distance $\Delta \bar{x}$ through the barrier: $\Delta
k_y=eB\Delta \bar{x}/\hbar$. Magnetic fields above 4~T separate
the occupied states of both systems in the $k_y$ space and
therefore tunneling at zero bias is no longer possible. In
Fig.~\ref{schematic} we have depicted the situation at finite
negative bias where the Fermi point $FP^{\parallel}$ matches the
dispersion curve of the lowest Landau level at the QH edge
resulting in a conductance peak. At even higher bias further peaks
are observed when $FP^{\parallel}$ touches the higher Landau
levels. Scanning both the voltage bias and the magnetic field we
can map out the entire $E^{\perp}$ vs. $k_y$ space using
$FP^{\parallel}$ as a probe for the Landau level dispersion
$E_n^{\perp}(k_y)$. The conductance maxima for the lowest two
Landau levels are indicated by little arrows in Fig.~\ref{cond}a
and their positions are plotted in the $V-B$ plane
(Fig.~\ref{cond}b) for comparison with model calculations. Spin
splitting is not resolved in these measurements.

\begin{figure}[h]
\begin{center}\leavevmode
\includegraphics[width=0.92\linewidth]{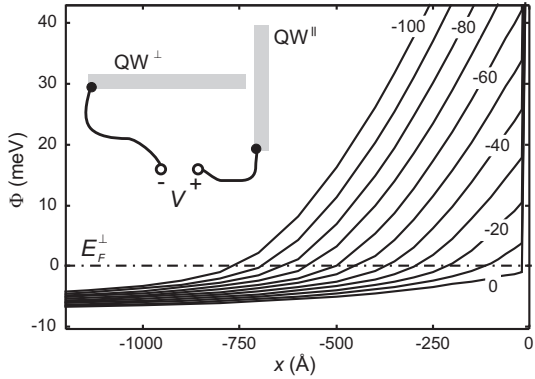}
\caption{Self-consistently calculated potential $\Phi (x,V)$ at
the edge of $QW^{\perp}$ for a series of applied bias voltages $V$
(denoted in mV) at $B=0$. The plotted lines represent the local
bottom of the 2D-subband in $QW^{\perp}$. Electronic states are
filled up to the Fermi level $E_F^{\perp}$. Note the onset of a
depletion region at 10~mV
bias.}\label{edgepot}\end{center}\end{figure}

From the measured $V-B$ relation we can deduce information about
the real space edge potential $\Phi (x,V)$ with B-field. For
comparison we have performed a self-consistent
Schroedinger-Poisson calculation of the edge potential at a biased
tunnel junction without magnetic field and plotted the result in
Fig.~\ref{edgepot}. There we have plotted $\Phi (x,V)$ as the
lowest 2D subband in $QW^{\perp}$ as a function of position $x$.
The subband is occupied up to the Fermi energy designated by the
dashed line. Even at zero bias the potential close to the edge is
not flat. Negative bias lifts the subband energy above the Fermi
level resulting in edge depletion of order 750~\AA\ at -100~mV,
for example.

We checked our self-consistent Schroedinger-Poisson calculations
against the analytical model of Levitov, et al. \cite{Levitov}
which was designed to model a similar device geometry. We find
agreement with the zero bias density distribution to within 5\% at
length scales of 400~\AA\ or more away from the barrier, but at
shorter distances and large biases the analytical model appears to
underestimate the edge depletion under negative bias as well as
the edge accumulation under positive bias, presumably a
consequence of the finite screening length in $QW^{\perp}$. The
potential distribution $\Phi (x,V)$ should be revealed in the
dispersion $E_n^{\perp}(k_y)$ of the QH edges when we apply a
magnetic field. Based on the results of the zero B-field
simulation for $\Phi (x,V)$ we have solved the Schroedinger
equation~(\ref{schroedinger2}) with $B$ field and calculated the
position of the expected maxima in the tunnel conductance. The
same simulation was done assuming the QH edge potential was
perfectly flat up to the barrier at any voltage bias. Both results
are plotted in Fig.~\ref{cond}. We observe significant deviation
of the experimental results from the flat band assumption
especially at large negative voltages, while the self-consistently
calculated edge potential shows excellent agreement with the
experiment.

\section{Conclusion}
With the presented measurement we are able to directly probe the
dispersion relation $E_n^{\perp}(k)$ of QH edge states. In this
paper we focused on the regime at high magnetic fields and
moderate negative bias where only the lowest Landau level is
occupied and the probed states are empty. We observe a shift in
the location of the conductance maxima that directly reveals the
band bending under bias at the tunnel junction, and is in
excellent agreement with a self-consistent model calculation.

\ack{We gratefully thank U. Z\"ulicke and M. Geller for helpful
discussions. This work was supported by Deutsche
Forschungsgemeinschaft via Schwerpunktprogramm
"Quanten-Hall-Systeme"}


\begin{thebibliography}{9}

\bibitem{CEO}
L. N. Pfeiffer, K. W. West, H. L. Stormer, J. P. Eisenstein, K. W.
Baldwin, D. Gershoni, and J. Spector, Appl. Phys. Lett. 56 (1990)
1697.

\bibitem{Chang PRL 96}
A. M. Chang, L. N. Pfeiffer, and K. W. West, Phys. Rev. Lett. 77
(1996) 2538.

\bibitem{Grayson PRL 98}
M. Grayson, D. C. Tsui, L. N. Pfeiffer, K. W. West, and A. M.
Chang, Phys. Rev. Lett. 80, 1062 (1998); Phys. Rev. Lett. 86
(2001) 2645.

\bibitem{Chang PRL 01}
A. M. Chang, M. K. Wu, C. C. Chi, L. N. Pfeiffer, and K. W. West,
Phys. Rev. Lett. 86 (2001) 143.

\bibitem{Hilke}
M. Hilke, D. C. Tsui, M. Grayson, L. N. Pfeiffer, and K. W. West,
Phys. Rev. Lett. 87 (2001) 186806.

\bibitem{Huber Physica E}
M. Huber, M. Grayson, M. Rother, R. A. Deutschmann, W. Biberacher,
W. Wegscheider, M. Bichler, and G. Abstreiter, Physica E 12 (2002)
125.

\bibitem{Zulicke}
U. Z\"ulicke, E. Shimshoni, and M. Governale, Phys. Rev. B 65
(2002) 241315.

\bibitem{Kang nature}
W. Kang, H. L. Stormer, L. N. Pfeiffer, K. W. Baldwin, and K. W.
West, Nature 403 (2000) 59.

\bibitem{Levitov}
L. S. Levitov, A. V. Shytov, and B. I. Halperin, Phys. Rev. B 64
(2001) 075322.


\end{thebibliography}
\end{document}